\documentstyle[12pt]{article} \topmargin -27pt \textwidth 6in \textheight
8.5in  \def\ka{\kappa}
 \def\f{\varphi} \def\x{\chi}
\def\la{\lambda} \def\ds{\zeta} \def\mf{m_{\f}} \def\mx{m_{\x}}
\def\d{\partial} \def\fo{\f_{0}} \def\zo{z_{1}} \def\zd{z_{2}}
\def\no{n_{1}} \def\af{A_{\no,2q}^{\f}} \def\ax{A_{\no,2q+1}^{\x}}
\def\s{\sigma} \def\e{\mbox{e}} 
\begin{document}

\begin{titlepage}
\title{Tree amplitudes at multiparticle threshold in a
model with softly broken $O(2)$ symmetry }
\author{ M.V. Libanov, V.A. Rubakov and S.V. Troitsky\\
 {\small{\em Institute for Nuclear Research of
the Russian Academy of Sciences, }}\\ {\small{\em 60th October Anniversary
prospect, 7a, Moscow 117312}}\\ }

\end{titlepage} \maketitle

\begin{center} {\bf Abstract} \end{center}

Tree amplitudes of the production of two kinds of scalar particles at
threshold from one virtual particle are calculated in a model of two scalar
fields with $O(2)$ symmetric quartic interaction and unequal masses. These
amplitudes exhibit interesting factorial and exponential behaviour at large
multiplicities. As a by-product we observe that the kinematically allowed
decay of one real particle into $n$ real particles of another kind, all at
rest, has zero tree amplitude in this model for $n>2$.

\newpage \section{Introduction} Recently, considerable attention have been
paid to the study of multiparticle production processes both in the
instanton sector \cite{Ringwald,Espinosa,Mattis,Peter} and in the
conventional perturbation theory \cite{Cornwall,Goldberg,Voloshin,Brown}.
In the latter case, most of the results has been obtained in simple
theories of one scalar field (or $p$ scalar fields with $O(p)$ symmetry),
where a number of exact expressions have been derived, e.g., for the
amplitudes of the production of $n$ particles at the multiparticle
threshold by one virtual particle, at the tree level
\cite{Voloshin,Brown,Kleiss} and in the one loop approximation
\cite{Voloshin',Smith,Kleiss'}. These amplitudes grow like $n!$ at large
$n$, and this non-trivial property is expected to persist to all orders of
the perturbation theory \cite{Voloshin'',Bubbles,Kleiss'}.

For calculating the tree amplitudes at the threshold, two methods have been
employed. One of them is based on recursion relations between diagrams with
different numbers of final particles \cite{Voloshin}, and the other makes
use of classical field equations with special boundary conditions
\cite{Brown}. Both of these methods have been generalized to incorporate
loop corrections \cite{Voloshin',Smith,Kleiss'} and both work nicely in the
simplest theories.

In this paper we study the tree amplitudes at the multiparticle threshold
in the theory of two real scalar fields with the lagrangian
\begin{equation} L=\frac{1}{2}(\d_{\mu}
\f)^{2}+\frac{1}{2}(\d_{\mu}\x)^{2}-\frac{m_{1}^{2}}{2}\f^{2}-
\frac{m_{2}^{2}}{2}
\x^{2}-
\la(\f^{2}+\x^{2})^{2} \label{2*} \end{equation} The interaction term is
$O(2)$-invariant, while the mass terms explicitly break $O(2)$ symmetry at
$m_{1}\ne m_{2}$. This model may be viewed as a testing ground for
different calculational techniques, and also has some features absent in
the simplest theory of one scalar field. For example, the model contains an
additional dimensionless parameter, $m_{1}/m_{2}$, and one can study the
dependence of the amplitudes on this parameter. Also, the class of the tree
amplitudes is wider than in the theory of one field because two kinds of
particles can be emitted. Finally, in the case of broken reflection
symmetry, $\f\to -\f$, there emerges a kinematically allowed possibility
(at $0>m_{2}^{2}>m_{1}^{2}$) of a decay of an {\em on-shell} $\f$-particle
at rest into $\x$-particles, all of which are also at rest (of course, an
interesting quantity is the amplitude of this decay and not the decay width
which is zero because of empty phase space).

We find that the classical solution method \cite{Brown} is convenient for
the evaluation of the tree amplitudes at the threshold in the model
(\ref{2*}). Due to a special symmetry of the corresponding classical
equations \cite{Ind}, the explicit solution of the relevant boundary value
problem can be found, and the explicit form of the tree amplitudes at
particle threshold is obtained in this paper for the processes $\f \to
n_{1}\f+n_{2}\x$ and $\x \to n_{1}\f+n_{2}\x$, where the initial particle
is off shell, for both unbroken ($<\f>=0$, $<\x>=0$) and broken ($<\f> \ne
0$, $<\x>=0$) cases. Not surprisingly, these amplitudes grow factorially at
large $n_{1}$ and $n_{2}$, but besides the expected factorials, the
amplitudes contain exponential terms depending non-trivially on $\mf/\mx$
and $\no/n_{2}$. Less obvious is the fact that the tree amplitudes of the
decay of an {\em on-shell} $\f$-particle at rest into $n_{2}$
$\x$-particles with zero momenta (kinematically allowed in the
spontaneously broken case) is zero for $n_{2}>2$.

In Sect.2 we obtain the general classical solution to the classical field
equations for space-independent fields. In Sect.3 we evaluate the tree
amplitudes at the threshold in the unbroken symmetry case $<\f>=<\x>=0$.
The broken symmetry case $<\f> \ne 0$, $<\x>=0$ is considered in Sect.4.
Sect.5 contains concluding remarks. \section{Solution to classical
equations} We consider the tree amplitudes of, say, the processes $\f\to
n_{1}\f+ n_{2}\x$, where the initial particle is off-shell and the final
particles are on-shell and have zero spatial momenta. According to
ref.\cite{Brown}, the generating function for these amplitudes,
$\f_{c}(z_{1},z_{2})$, is the solution to the classical field equations for
space-independent fields, obeying the condition that \begin{equation} \f
\to z_{1}+\f_{0} \mbox{,\hspace{1cm}  } \x \to  z_{2} \label{6+}
\end{equation} as $\la\to 0$, where \begin{equation}
z_{1}=a_{1}e^{im_{\f}t} \mbox{,\hspace{1cm}  } z_{2}=a_{2}e^{im_{\x}t}
\label{6*} \end{equation} $m_{\f}$ and $m_{\x}$ are the particle masses,
$\f_{0}=<\f>$ (hereafter we assume $<\x>=0)$, and $a_{1}$ and $a_{2}$ are
arbitrary constants. The amplitudes are then \begin{equation}
A_{n_{1},n_{2}}=\frac{\d^{\no}}{\d\zo^{\no}}\:\frac{\d^{n_{2}}}{\d\zd^{n_{2}}}
\:\f_{c}(z_{1},z_{2})\Big|_{\zo=\zd=0}
\label{6**} \end{equation}

Thus, we have to consider the hamiltonian for space-independent fields (we
set the spatial volume equal to 1; the dependence on the volume can be
easily restored),
\[H=\frac{1}{2}(\dot{\f})^{2}+\frac{1}{2}(\dot{\x})^{2}+\frac{m_{1}^{2}}{2}
\f^{2}+\frac{m_{2}^{2}}{2}\x^{2}+
\la(\f^{2}+\x^{2})^{2} \] It has been found in ref.\cite{Ind} that this
hamiltonian possesses a non-trivial symmetry and thus is integrable. It has
been also pointed out \cite{Ind'} that the system is separable in elliptic
coordinates. So, we introduce new variables $\xi(t)$ and $\eta(t)$, \[\f=\s
\sqrt{(\xi^{2}-1)(1-\eta^{2})},\] \[\x=\s \xi \eta, \;\;\;\;\;\;\;\;
\s=\mbox{const}.\] By choosing \[\s^{2}=\frac{m_{1}^{2}-m_{2}^{2}}{2\la}\]
we find that the variables indeed separate in the Hamilton-Jacobi equation
and we find for the truncated action $S(\xi,\eta)$,

\begin{eqnarray} (\xi^{2}-1)\Big(\frac{\d S}
{\d\xi}\Big)^{2}+2\s^{2}\Big\{\la\s^{4}\xi^{6}+\Big(\frac{m_{1}^{2}}{2}
-2\la\s^{2}\Big)\s^{2}\xi^{4}-\Big(\frac{m_{1}^{2}}{2}\s^{2}-\la\s^{4}+E\Big)
\xi^{2}\Big\}=
\nonumber \\ (\eta^{2}-1)\Big(\frac{\d
S}{\d\eta}\Big)^{2}+2\s^{2}\Big\{\la\s^{4}\eta^{6}+\Big(\frac{m_{1}^{2}}{2}
-2\la\s^{2}\Big)\s^{2}\eta^{4}-\Big(\frac{m_{1}^{2}}{2}\s^{2}-\la\s^{4}+E\Big)
\eta^{2}\Big\}=
\nonumber \\ 2\s^{2}J \label{8*} \end{eqnarray} where $E$  is the classical
energy and $J$ is another integral of motion. Therefore, the truncated
action is \[S=F(\xi)+F(\eta),\] where $$ F(\xi)=\int\! d\xi\,
\sqrt{\frac{2\s^{2}J-2\s^{2}\xi^{2}[\la\s^{4}\xi^{4}+(m_{1}^{2}/2
-2\la\s^{2})\s^{2}\xi^{2}-(m_{1}^{2}\s^{2}/2-\la\s^{4}+E)]}{\xi^{2}-1}}.$$
The action function is then \[S_{tot}=-Et+F(\xi)+F(\eta).\] So, the general
solution is determined by \[\frac{\d S_{tot}}{\d E}\equiv -t+\frac{\d
F(\xi)}{\d E}+\frac{\d F(\eta)}{\d E}=-t_{0},\] \[\frac{\d S_{tot}}{\d
J}=\beta,\] where $\beta$ and $t_{0}$ are arbitrary constants. Thus, the
general solution is finally obtained in the implicit form \begin{equation}
\s^{2}\int\! d \xi\, \frac{\xi^{2}}{g(\xi)}+\s^{2} \int\! d\eta\,
\frac{\eta^{2}}{g(\eta)}= t-t_{0}, \label{9*} \end{equation}
\begin{equation} \s^{2}\int\! d \xi\, \frac{1}{g(\xi)}+\s^{2} \int\!
d\eta\, \frac{1}{g(\eta)}=\beta, \label{9**} \end{equation} where
\begin{equation}
g(\xi)=\sqrt{\xi^{2}-1}\sqrt{2\s^{2}J-2\s^{2}\xi^{2}\Big[\la\s^{4}\xi^{4}+
\Big(\frac{m_{1}^{2}}{2}
-2\la\s^{2}\Big)\s^{2}\xi^{2}-\Big(\frac{m_{1}^{2}}{2}
\s^{2}-\la\s^{4}+E\Big)\Big]} \label{9+} \end{equation} Since we are
interested in complex solutions, the arbitrary constants $E$, $J$, $\beta$
and $t_{0}$ are also, in general, complex. We will see, however, that in
both broken and unbroken symmetry cases, the relevant solutions correspond
to $J=0$, \mbox{$E=H(<\f>)=$} \mbox{(vacuum energy)}, while $t_{0}$ and
$\beta$ are determined by the conditions (\ref{6+}). \section{Unbroken
reflection symmetry} Let us first consider the case $m_{1}^{2}>0$,
$m_{2}^{2}>0$ when $<\f>\equiv\f_{0}=0$. To ensure the conditions
(\ref{6+}) we note that in the limit $\la \to 0$, the fields $\f$ and $\x$
do not interact with each other, and their energies, $E_{\f}$ and $E_{\x}$,
are the two integrals of motion. So, the integrals of motion of the
non-linear problem, $E$ and $J$, may be expressed through $E_{\f}$ and
$E_{\x}$ in this limit. Obviously, $E=E_{\f}+E_{\x}$, and it is
straightforward to see from eq.(\ref{8*}) that $J=-E_{\x}$ at $\la \to0$.

For positive frequency solutions, eqs. (\ref{6+}) and (\ref{6*}), one has
$E_{\f}=E_{\x}=0$, so the relevant solution to the non-linear equations is
determined by eqs. (\ref{9*}), (\ref{9**}) and (\ref{9+}) with \[E=J=0\]
Then the system (\ref{9*}), (\ref{9**}) can be solved explicitly, and the
solution obeying eqs. (\ref{6+}) and (\ref{6*}) is (see Appendix A for
details) \begin{equation} \f=z_{1}\bigl(1-\la
\frac{\ka}{2m_{2}^{2}}z_{2}^{2}\bigr)\Bigl(1-\frac{\la}{2m_{1}^{2}}z_{1}^{2}-
\frac{\la}{2m_{2}^{2}}z_{2}^{2}+\la^2
\frac{\ka^{2}}{4m_{1}^{2}m_{2}^{2}}z_{1}^{2}z_{2}^{2}\Bigr)^{-1}
\label{12*} \end{equation} \begin{equation} \x=z_{2}\bigl(1+\la
\frac{\ka}{2m_{1}^{2}}z_{1}^{2}\bigr)\Bigl(1-\frac{\la}{2m_{1}^{2}}z_{1}^{2}-
\frac{\la}{2m_{2}^{2}}z_{2}^{2}+\la^2
\frac{\ka^{2}}{4m_{1}^{2}m_{2}^{2}}z_{1}^{2}z_{2}^{2}\Bigr)^{-1}
\label{12**} \end{equation} where \[\ka=\frac{m_{1}-m_{2}}{m_{1}+m_{2}}\]
Notice that at $z_{2}=0$, the function (\ref{12*}) coincides with the known
solution for one scalar field \cite{Brown}, as it should.

The amplitudes $\f \to n_{1}\f+n_{2}\x$ are given by eq. (\ref{6**}). They
are non-zero at \[n_{1}=2p+1,\] \[n_{2}=2q,\] where $p$ and $q$ are
integer. We find \begin{equation}
A_{p,q}=\frac{(2p+1)!\,(2q)!}{m_{1}^{2p}m_{2}^{2q}}\Bigl(\frac{\la}{2}\Bigr)
^{p+q}
\sum_{l=0}^{\min{(p,q)}} (-1)^{l} \ka^{2l}
\frac{(p+q-l-1)!}{l!\,(p-l)!\,(q-l)!} [p+q-\ka q-(1-\ka)l] \label{13*}
\end{equation} The finite sum in eq.(\ref{13*}) can be transformed into the
Jacobi polynomials of the argument $(1-2\ka^{2})$ (see Appendix B), so that
the expression for the amplitude  takes the form \[
A_{p,q}=\frac{(2p+1)!\,(2q)!}{m_{1}^{2p}m_{2}^{2q}}\Bigl(\frac{\la}{2}\Bigr)
^{p+q}
(-1)^{p}\frac{p+q-1}{pq}\times \] \begin{equation} \Bigl[(p+q-\ka
q)P_{p-1}^{(1-q-p,\, 0)}(1-2\ka^{2})-(q-1)(1-\ka)\ka^{2} P_{p-2}^{(2-p-q,\,
1)}(1-2\ka^{2})\Bigr] \label{14+} \end{equation} An equivalent form of the
amplitude can be obtained by transforming the sum in eq.(\ref{13*}) into
the hypergeometric function F (see Appendix B). At $ q\geq p$ we have \[
A_{p,q}=\frac{(2p+1)!\,(2q)!}{m_{1}^{2p}m_{2}^{2q}}\Bigl(\frac{\la}{2}\Bigr)
^{p+q}
\times \] \begin{equation}
\Bigl\{\ka^{2p}(-1)^{p}\frac{(q-1)!}{p!\,(q-p)!}\bigl[pF(-p,q;q-p+1;\frac{1}
{\ka^{2}})+
(q-p)(1-\ka)F(-p,q;q-p;\frac{1}{\ka^{2}})\bigr]\Bigl\} \label{14*}
\end{equation} while the expression at $q<p$ is obtained by interchanging
$p\leftrightarrow q$ in the curly brackets in eq. (\ref{14*}).

Eq.(\ref{13*}) (or, equivalently, eqs.(\ref{14+}) or (\ref{14*})) is the
exact formula for the tree amplitudes. Its asymptotics at large $p$ and $q$
and $p/q= \mbox{fixed}$ has the form \begin{equation} A_{p,q}\propto
\frac{(2p+1)!\,(2q)!}{m_{1}^{2p}m_{2}^{2q}} \exp\bigl[(p+q)G\bigl(
\frac{m_{1}}{m_{2}};\frac{p}{q}\bigr)\bigr] \label{15*} \end{equation} The
explicit expression, given in Appendix C, is not very illuminating. Notice
that $A_{p,q}$ grows factorially as could have been expected.

A particularly simple asymptotics emerges at $p\to\infty$, $q\to\infty$,
$p/q\to 1$. In that case we obtain (see Appendix C) \begin{equation}
A_{p,q}\propto
\frac{(2p+1)!\,(2q)!}{m_{1}^{2p}m_{2}^{2q}}\Bigl(\frac{\la}{2}\Bigr)^{2p}
\Bigl[\frac{(\sqrt{m_{1}}+\sqrt{m_{2}})^{2}}{m_{1}+m_{2}}\Bigr]^{2p}C(p)
\label{15+} \end{equation} where $C(p)$ depends also on $m_{1}/m_{2}$ and
grows like some power of $p$.

\section{Broken reflection symmetry}

At $m_{1}^{2}<0$, $\,m_{2}^{2}>m_{1}^{2}$, the symmetry $\f\to -\f$ is
spontaneously broken, and \[ <\f>=\f_{0}=\frac{|m_{1}|}{2\sqrt\la} \] The
masses of excitations around this vacuum are \begin{equation}
m_{\f}=\sqrt{2}|m_{1}|\mbox{,\hspace{1cm}
}\mx=\sqrt{|m_{1}|^{2}+m_{2}^{2}} \label{16**} \end{equation}

The conditions (\ref{6+}) and (\ref{6*}) are satisfied when the integrals
of motion are equal to \[ J=0\mbox{,\hspace{1cm}
}E=-\frac{m_{1}^{4}}{16\la}=E_{vac} \] The explicit solution to the
equations of motion, that obeys the conditions (\ref{6+}) and (\ref{6*}),
is obtained in the same way as in Sect.3. We find \[
\f=\fo\Bigl(1+\frac{\zo}{
2\fo}+\frac{2\la}{4\mx^{2}-\mf^{2}}\zd^{2}+\frac{\la}
{\fo}\frac{2\mx-\mf}{(2\mx+\mf)^{3}}\zo\zd^{2}\Bigr)\times \]
\begin{equation}
\Bigl(1-\frac{\zo}{2\fo}-\frac{2\la}{4\mx^{2}-\mf^{2}}\zd^{2}+\frac{\la}
{\fo}\frac{2\mx-\mf}{(2\mx+\mf)^{3}}\zo\zd^{2}\Bigr)^{-1}, \label{16*}
\end{equation} \hspace{0.7cm} \[
\x=\zd\Bigl(1-\bigl(\frac{2\mx-\mf}{2\mx+\mf}\Bigr)\frac{\zo}{2\fo}\Bigr)\times
\] \begin{equation}
\Bigl(1-\frac{\zo}{2\fo}-\frac{2\la}{4\mx^{2}-\mf^{2}}\zd^{2}+\frac{\la}
{\fo}\frac{2\mx-\mf}{(2\mx+\mf)^{3}}\zo\zd^{2}\Bigr)^{-1} \label{16+}
\end{equation}

The field (\ref{16*}) reduces to the known solution for spontaneously
broken theory of one scalar field \cite{Brown} at $\zd=0$. On the other
hand, eq.(\ref{16+}) does not coincide at $ \zo=0$ with the solution for
the theory of one field with unbroken symmetry \cite{Brown}; in the
diagrammatic language this corresponds to the existence of the diagrams of
fig.1b absent in the theory of one field.

In the case of broken symmetry we should distinguish between the processes
$\f\to n_{1}\f+n_{2}\x$ and $\x\to n_{1}\f+n_{2}\x$. The amplitudes of the
former, which we denote by $A_{\no,2q}^{\f}$ (the number of $\x$-particles
in the final state is even), are generated by $\f(\zo,\zd)$. The amplitudes
of the latter, $\ax$, are generated by $\x(\zo,\zd)$. We find for the
process $\f\to n_{1}\f+2q\x$, \[
\af=\frac{\no!(2q)!}{2\mf^{\no-1}(4\mx^{2}-\mf^{2})^{q}}(\sqrt{2\la})^{2q+
\no-1}\times
\] \begin{equation} \sum_{l=0}^{\min{(\no,q)}} (-1)^{l} \ds^{2l}
\frac{(\no+q-l-2)!}{l!\,(\no-l)!\,(q-l)!}
\bigl[(\no+q-l-1)(2\no+2q-3l)+(\no-l)(q-l)\ds^{2}\bigr] \label{17*}
\end{equation} where \[ \ds=\frac{2\mx-\mf}{2\mx+\mf} \]

In analogy to Sect.3, the amplitudes in eq.(\ref{17*}) can  be expressed
through the Jacobi polynomials or hypergeometric function. We present here
the latter form only, which reads \[
\af=\frac{\no!(2q)!}{\mf^{\no-1}(4\mx^{2}-\mf^{2})^{q}}(\sqrt{2\la})^{2q+\no-1}
\times
\] \begin{equation}
\Bigl\{(-1)^{\no}\ds^{2\no}\frac{q!}{(q-\no)!\no!}\bigl[F(-\no,q+1;q-\no+1;
\frac{1}{\ds^{2}})
-\frac{\no}{q}F(-\no+1,q;q-\no+1;\frac{1}{\ds^{2}})\bigr]\Bigr\}
\end{equation} Here it is assumed that $q\geq\no$; at $q<\no$ the
expression for the amplitude is obtained by substituting
$q\leftrightarrow\no$ in curly brackets.

{}From the latter expression and eq.(\ref{C3*}), the asymptotics of the
amplitude at \mbox{ $\no\to\infty$}, $q\to\infty$, $\no/q\to 1$ can be
obtained, \begin{equation}
\af\propto\frac{\no!(2q)!}{\mf^{\no-1}(4\mx^{2}-\mf^{2})^{q}}(\sqrt{2\la})^{2q+
\no-1}
\Bigl[\frac{(\sqrt{2\mx}+\sqrt{\mf})^{2}}{2\mx+\mf}\Bigr]^{2\no}\;C(\no,q)
\label{19*} \end{equation} where $C(\no,q)$ has power-like behaviour in
this limit. The asymptotics in the general regime, $\no\to\infty$,
$q\to\infty$, $\no/q=$fixed, can also be found and has the general form of
eq.(\ref{15*}).

{}From eq.(\ref{16+}) we find the amplitudes of the processes
$\x\to\no\f+(2q+1)\x$, \newpage  \[
\ax=\frac{\no!(2q+1)!}{\mf^{\no}(4\mx^{2}-\mf^{2})^{q}}(\sqrt{2\la})^{2q+\no}
\times
 \] \[ \sum_{l=0}^{\min{(\no,q)}} (-1)^{l} \ds^{2l}
\frac{(\no+q-l-1)!}{l!\,(\no-l)!\,(q-l)!} \bigl[\no+q-l-\ds(\no-l)\bigr] \]
 This expression can again be rewritten in terms
of the Jacobi polynomials or hypergeometric function. At $\no\to\infty$,
$q\to\infty$, $\no/q\to 1$ we have the following asymptotics,
\begin{equation}
\ax\propto\frac{\no!(2q+1)!}{\mf^{\no}(4\mx^{2}-\mf^{2})^{q}}(\sqrt{2\la})
^{2q+\no}
\Bigl[\frac{(\sqrt{2\mx }+\sqrt{\mf})^{2}}{2\mx+\mf}\Bigr]^{2\no}\;
C(\no,q) \label{20*} \end{equation} We observe that both $\ax$ and $\af$
increase factorially at large number of final particles.

To conclude this section, we point out that at certain values of
$|m_{1}/m_{2}|$ such that $m_{\f}=n_{2}m_{\x}=2qm_{\x}$, there exists a
kinematically allowed possibility of the decay of a $\f$-particle at rest
into $n_{2}=2q$ $\x$-particles, all of which are also at rest (according to
eq.(\ref{16**}), this requires $0>m_{2}^{2}>m_{1}^{2}$). Eq.(\ref{17*})
tells us that this amplitude is in fact zero at the tree level, unless
$n_{2}=2$ ($q=1$). Indeed, the amplitude $A_{0,2q}^{\f}$ corresponds to the
diagrams like those shown in fig.2, where {\em the propagator of the
initial particle is included}. So, the decay amplitude $A_{\f\to
2q\x}^{decay}$ is obtained from $A_{0,2q}^{\f}$ by truncating the leg
corresponding to the initial particle, \begin{equation} A_{\f\to
2q\x}^{decay}=(4q^{2}m_{\x}^{2}-\mf^{2})A_{0,2q}^{\f} \label{20a*}
\end{equation} At $q=1$, the amplitude of eq.(\ref{17*}) has a pole at
$\mf^{2}=4q^{2}\mx^{2}=4\mx^{2}$, and one recovers the obvious result, \[
A_{\f\to 2\x}^{decay}=- 2 i \sqrt{2\la}\mf \] On the other hand, at $q>1$
($n_{2}>2$), the amplitude of eq.(\ref{17*}) does not have a pole at
$\mf^{2}=4q^{2}\mx^{2}$, and the decay amplitude (\ref{20a*}) vanishes.
This means, for example, that the diagrams of fig.2 with truncated initial
legs, cancel each other for on-shell initial  and final particles, all at
rest (for the decay $\f\to 4\x$ this fact can of course be checked by the
direct calculation of the diagrams).

\section{Conclusion} The model of two scalar fields with $O(2)$ symmetric
interaction and unequal masses has the property of integrability and
separability of its classical equations for spatially homogeneous fields.
This property enabled us to obtain the explicit formulas for the tree
amplitudes at particle threshold for both unbroken and broken reflection
symmetry $\f\to-\f$. These amplitudes are expressed through the Jacobi
polynomials or, equivalently, finite hypergeometric series.

The behaviour of the tree amplitudes of the creation of $n_{1}$
$\f$-particles and $n_{2}$ $\x$-particles at threshold at large $n_{1}$,
$n_{2}$ and fixed $n_{1}/n_{2}$ is suggestive. It has the factorial growth
and is exponential in the free parameters $m_{\f}/m_{\x}$ and
$n_{1}/n_{2}$, \begin{equation} A_{n_{1},n_{2}}\propto
n_{1}!\,n_{2}!\,\la^{(n_{1}+n_{2})/2}
\exp\bigl[(n_{1}+n_{2})G\bigl(\frac{n_{1}}{n_{2}},\frac{m_{\f}}{m_{\x}}\bigr)
\bigr]
\label{22*} \end{equation} up to well defined powers of $m_{1}$ and $m_{2}$
and pre-exponential factors growing power-like at large $n_{1}$, $n_{2}$.
The function $G\bigl(\frac{n_{1}}{n_{2}},\frac{m_{\f}}{m_{\x}}\bigr)$ is
calculable at the tree level; its explicit form at arbitrary $n_{1}/n_{2}$
in the unbroken case can be read off from the eq. (\ref{C2**}), while at
$n_{1}/n_{2}\to 1$ the corresponding expressions are particularly simple,
see eqs. (\ref{15+}), (\ref{19*}) and (\ref{20*}).

Eq. (\ref{22*}) indicates that there may exist a semiclassical-type
procedure for calculating the leading exponential behaviour of the
amplitudes at large $n_{1}$, $n_{2}$. Such a procedure is explicit in the
recent study \cite{Bubbles} of the amplitudes in the theory of one scalar
field with broken symmetry. So, one may anticipate that this feature is
characteristic to a wide class of bosonic theories.

A peculiar property of the model with $<\f>\neq 0$ is that the decay of a
$\f$-particle into $2q$ $\x$-particles at rest, although kinematically
allowed at appropriate values of particle masses, has zero tree amplitude
for $2q>2$. This fact may or may not be related to the integrability of the
classical theory of spatially homogeneous fields. We hope to discuss the
cancellation of the corresponding diagrams (that occurs also for some other
processes) in future.

The authors are indebted to D.T. Son, E.E. Tareyeva and P.G. Tinyakov for
helpful discussions. One of us (V.R.) thanks M.B. Voloshin and L. McLerran
for stimulating conversations at the initial stage of this work. The work
of M.L. and S.T. is supported in part by the Weingart Foundation through a
cooperative agreement with the Department of Physics at UCLA.
\section*{Appendix A. Classical solution} The integrals in eqs.(\ref{9*}),
(\ref{9**}) are straightforward to evaluate at $E=J=0$. The system
(\ref{9*}), (\ref{9**}) then takes the form \[
\frac{(\psi+m_{1})(\mu+m_{1})}{(\psi-m_{1})(\mu-m_{1})}=
\exp[\mp2im_{1}(t-t_{0})]
\] \begin{equation}
\frac{(\psi+m_{2})(\mu+m_{2})}{(\psi-m_{2})(\mu-m_{2})}=\exp[\mp2im_{2}(t-t_{0}-
  2\beta\s^{2})] \label{A1*} \end{equation} where
$\psi=\sqrt{a\xi^{2}+m_{2}^{2}}$,  $\mu=\sqrt{a\eta^{2}+m_{2}^{2}}$,
$a=m_{1}^{2}-m_{2}^{2}$. We note that the original fields $\f$ and $\x$ can
be expressed in terms of $\mu$ and $\psi$,
\[\f^{2}=\frac{1}{2a\la}(\psi^{2}-m_{1}^{2})(m_{1}^{2}-\mu^{2})\]
\[\x^{2}=\frac{1}{2a\la}(\psi^{2}-m_{2}^{2})(\mu^{2}-m_{2}^{2})\]
Therefore, to find the fields $\f$ and $\x$ one does not need to evaluate
$\xi (t)$ and $\eta (t)$, but one just has to solve the algebraic equations
(\ref{A1*}). This is easily done in terms of variables $(\psi \mu)$ and
$(\psi+\mu)$. We find that the correct behaviour (\ref{6+}),(\ref{6*})
corresponds to the upper sign in eq. (\ref{A1*}), and that the solution is
given by eqs. (\ref{12*}), (\ref{12**}).

\section*{Appendix B. Reduction to Jacobi polynomials and hypergeometric
functions} The sum in eq.(\ref{13*}) can be rewritten in the following
form, \[ \sum=\sum_{l} \frac{(p+q-1)!}{p!\,q!}\bigl[(p+q-\ka
q)-(1-\ka)l\bigr]\frac{\ka^{2l}}{l!} \times\] \[\frac{\bigl[(-p+1)\cdots
(-p+l)\bigr]\bigl[(-q+1)\cdots (-q+l)\bigr]} {\bigl[-(p+q-1)+1\bigr] \cdots
\bigl[-(p+q-1)+l \bigr] }  \] This should be compared with the
hypergeometric series \cite{Gradstein,Abramowitz}
\[F(a,b;c;z)=\frac{\Gamma(c)}{\Gamma(a)\Gamma(b)} \sum_{l}
\frac{\Gamma(a+l)\Gamma(b+l)}{\Gamma(c+l)} \: \frac{z^{l}}{l!}  \] The
direct comparison gives \[ \sum =\frac{(p+q-1)!}{p!\,q!} \bigl[(p+q-\ka q)
F(-p+1,-q+1;-p-q+2;\ka^{2})- \] \[
(1-\ka)\ka^{2}\frac{(-p+1)(-q+1)}{(-p-q+2)}
F(-p+2,-q+2;-p-q+3;\ka^{2})\bigr] \] Eq. (\ref{14+}) is then obtained by
making use of the relations \cite{Gradstein,Abramowitz} between the Jacobi
polynomials and hypergeometric functions.

The representation (\ref{14*}) is obtained by changing the variable in the
sum, $l\to p-l$, and then by using essentially the same trick.

\section*{Appendix C. Asymptotics of the amplitudes} Let us first evaluate
the asymptotics of the Jacobi polynomials $P^{(\alpha,\beta)}_{r}(x)$ at
$r\to\infty$, $\alpha\to -\infty$, $\alpha/r=-\theta=\mbox{const}$. We make
use of the generating function for the Jacobi polynomials, \[{\cal
F}(y)=2^{\alpha+\beta}R^{-1}(1-y+R)^{-\alpha}(1+y+R)^{-\beta}\] where
$R=\sqrt{1-2xy+y^{2}}$, $|y|<1$. The Jacobi polynomials are
\begin{equation} P^{(\alpha,\beta)}_{r}(x)=\frac{d^{r}\!{\cal
F}}{dy^{r}}\Big|_{y=0}= \frac{1}{r!} \int\! \bar{y}^{r}\e^{-y\bar{y}} {\cal
F}(y) \,dy\,d\bar{y} \label{C1*} \end{equation} The integral in this
equation is evaluated in the saddle point approximation. The saddle point
is \[\bar{y}=\frac{r}{y},\] \begin{equation} y=\frac{1}{4(\theta-1)}\:
\Bigl[-x(\theta-2)^{2}-\theta^{2}+(\theta-2)^{2}
\sqrt{(x+1)(x+1+\frac{8(\theta-1)}{(\theta-2)^{2}})}\;\Bigr] \label{C2*}
\end{equation} In this way we obtain the asymptotics \begin{equation}
P^{(\alpha,\beta)}_{r}(x)=2^{\alpha}y^{r}(1-y+R)^{\alpha}\,C(r,\alpha)
\label{C2+} \end{equation} where $y$ and $R(y)$ are taken at the saddle
point (\ref{C2*}), and $C(r,\alpha)$ behaves like some power of $r$.
Inserting eq. (\ref{C2+}) into eq. (\ref{14+}) we obtain the asymptotics of
the amplitude at \[p\to\infty \mbox{, \hspace{1cm}} q\to\infty,\]
\[\theta=1+\frac{q}{p}=\mbox{fixed},\] \[x=1-2\ka^{2}=\mbox{fixed}\] in the
following form \begin{equation} A_{p,q}=(-1)^{p}\,
\frac{(2p+1)!\,(2q)!}{m_{1}^{2p}m_{2}^{2q}}
q\Bigl[\frac{\la(yq/p+1)}{8\ka^{2}(q/p-1)}\Bigr]^{q+p} \frac{1}{y^{q}}\,
C(p,q) \label{C2**} \end{equation} where $C(p,q)$ behaves like a power of
$p$ and $q$, \  $y(\theta,x)$ is given by eq. (\ref{C2*}). Obviously, this
expression has the form of eq. (\ref{15*}).

Eq. (\ref{C2**}) looks singular at $p/q \to 1$, i.e. $\theta=2$. However,
the singularity is, in fact, absent because $y=-1$ at that point. The
asymptotics of the amplitude in this regime can be obtained either by the
study of the integral in eq.(\ref{C1*}) or by making use of eq. (\ref{14*})
and the appropriate asymptotics of the hypergeometric function. Namely, at
$\Lambda\to\infty$ one has \cite{BE}

\begin{equation} F(\Lambda,-b-\Lambda;c;
\frac{1}{\ka^{2}})=\frac{\Gamma(1+b+\Lambda)}{\Gamma(c+b+\Lambda)}\:
\e^{\Lambda\xi}\: C(\Lambda) \label{C3*} \end{equation} where
\[\e^{\xi}=\frac{\ka^{2}-2-2\sqrt{1-\ka^{2}}}{\ka^{2}}\] and $C(\Lambda)$
behaves as some power of $\Lambda$ at $\Lambda\to\infty$. Eqs. (\ref{C3*})
and (\ref{14*}) lead to \mbox{eq. (\ref{15+})} in a straightforward way.

In this Appendix we did not discuss the pre-exponential factors in
equations like eq. (\ref{C2+}) or eq. (\ref{C3*}). In principle, there
could have occurred cancellations between different terms in eq.
(\ref{14+}) or (\ref{14*}). We have checked by calculating the
pre-exponentials that the cancellations occur in none of these formulas,
i.e., the exponential behaviour, eqs. (\ref{15*}), (\ref{15+}) indeed takes
place.

 \begin{figure} \thicklines
\begin{picture}(432,120)

\put(250,100){\line(1,0){50}} \multiput(300,100)(15,10){3}{\line(3,2){10}}
\put(341,127){\line(3,-2){40}} \put(341,127){\line(3,2){40}}
\put(300,100){\line(3,-2){80}} \put(300,0){Fig.1b}
\put(0,100){\line(1,0){50}} \put(50,100){\line(3,2){70}}
\put(50,100){\line(1,0){70}} \put(50,100){\line(3,-2){70}}
\put(50,0){Fig.1a} \end{picture} \caption{$\x\to3\x$ diagrams in the model
with broken reflection symmetry $\f\to -\f$; solid and dashed lines
correspond to the fields $\x$ and $\f$, respectively.}
\begin{picture}(432,350) \multiput(0,100)(15,0){6}{\line(1,0){10}}
\put(40,100){\line(3,2){30}} \put(40,100){\line(3,-2){30}}
\put(85,100){\line(3,2){30}} \put(85,100){\line(3,-2){30}}
\multiput(240,100)(15,0){3}{\line(1,0){10}}
\multiput(280,100)(15,10){3}{\line(3,2){10}}
\multiput(280,100)(15,-10){3}{\line(3,-2){10}}
\put(320,127){\line(3,2){30}} \put(320,127){\line(3,-2){30}}
\put(320,73){\line(3,2){30}} \put(320,73){\line(3,-2){30}}
\multiput(0,250)(15,0){3}{\line(1,0){10}} \put(40,250){\line(3,-2){60}}
\put(40,250){\line(3,2){60}} \put(70,270){\line(1,0){30}}
\put(70,270){\line(3,2){30}} \put(70,270){\line(3,-2){30}}
\multiput(240,250)(15,0){3}{\line(1,0){10}}
\multiput(310,270)(15,10){3}{\line(3,2){10}} \put(280,250){\line(3,2){30}}
\put(280,250){\line(3,-2){90}} \put(310,270){\line(3,-2){30}}
\put(350,297){\line(3,2){30}} \put(350,297){\line(3,-2){30}}
\put(185,0){Fig.2} \end{picture} \caption{Diagrams for the decay $\f\to
4\x$ in the model with broken reflection symmetry.} \end{figure}
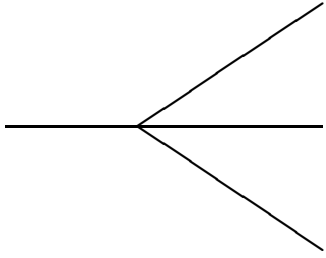
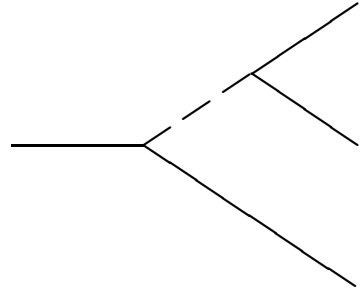
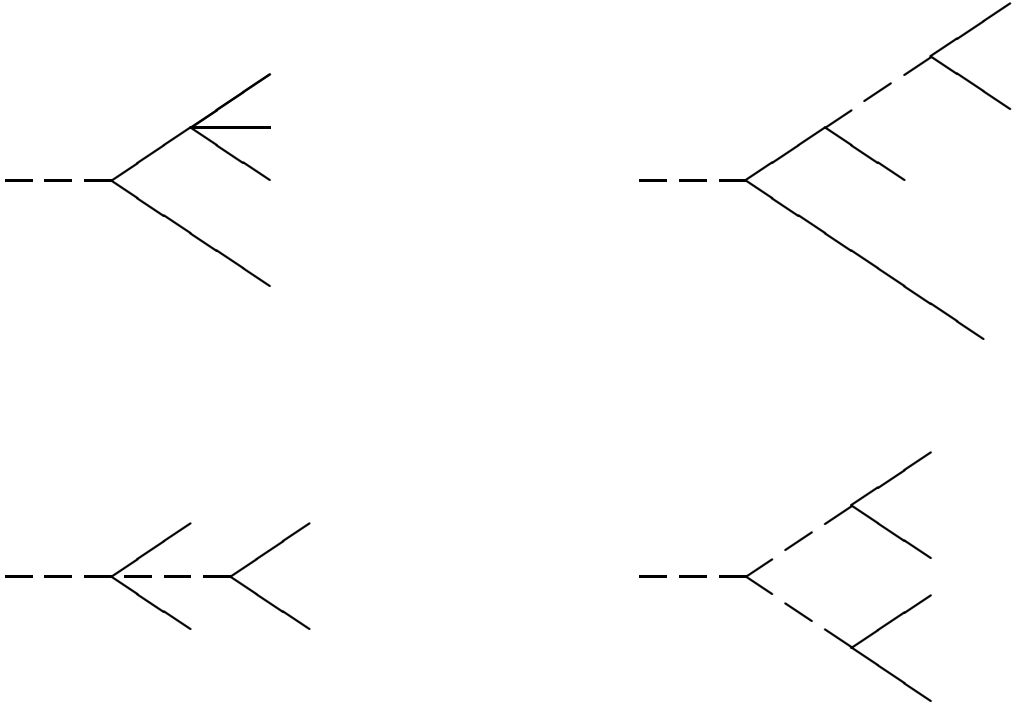
\end{document}